\documentclass[12pt]{article} 

\usepackage[utf8]{inputenc} 

\usepackage{geometry} 
\usepackage{graphicx, subfig} 

\usepackage{booktabs} 
\usepackage{array} 

\usepackage{fancyhdr} 
\usepackage{cite}

\usepackage{amsmath, amsfonts, amsthm, amssymb, amscd}
\usepackage{indentfirst}

\newtheorem{theorem}{Theorem}

\title{Perfect Information Hearthstone is PSPACE-hard}
\author{Zhujun Zhang \thanks{e-mail: zhangzhujun1988@163.com} \\Government Data Management Center of \\Fengxian District, Shanghai, China}

\date{May 22, 2023} 

\begin{document}
\maketitle

\begin{abstract}

We consider the computational complexity of Hearthstone which is a popular online CCG (collectible card game).
We reduce a PSPACE-complete problem, the partition game, to perfect information Hearthstone in which there is no hidden information or random elements.
In the reduction, each turn in Hearthstone is used to simulate one choice in the partition game.
It is proved that determining whether the player has a forced win in perfect information Hearthstone is PSPACE-hard.

\end{abstract}

\section{Introduction}

\textit{Hearthstone: Heroes of Warcraft} is a popular online CCG (collectible card game) developed by Blizzard Entertainment.
In the game of Hearthstone, there are three game types, Constructed, Battlegrounds, and Mercenaries type, and in this paper we just consider the constructed type which is the main and classic type.
In the constructed type, two players use virtual cards to construct their decks.
Players take actions on their own turns, including summoning and attacking with minions, casting spells, and controlling their heroes.
Each player's object is to reduce the health of the opponent's hero to zero.

In the last decades, the computational complexity of many video games was determined.
In 2000, Kaye \cite{minesweeper} proved Minesweeper, which is a classic game on the Windows system, to be NP-complete.
In 2004, Breukelaar, Demaine, and Hohenberger et al. \cite{tetris} proved the NP-hardness of the famous video game Tetris.
In 2015, Viglietta \cite{lemmings} proved Lemmings to be PSPACE-complete.
Aloupis, Demaine, and Guo et al. \cite{nintendogames} researched the computational complexity of some classic Nintendo games.
In 2016, Demaine, Viglietta, and Williams \cite{mariopspace} proved Super Mario to be PSPACE-complete.
Researchers also introduced some hardness frameworks, such as \cite{nintendogames} \cite{gamingishardjob}.
With the help of frameworks, people could prove the hardness of games relatively easily.
Other results about the computational complexity of games and puzzles could be found in Hearn and Demaine’s summary book \cite{gamespuzzlescomputation}.

Recently the computational complexity of card games was considered.
In 2017, Bosboom and Hoffmann \cite{netrunnernphard} discussed the complexity of Netrunner, which is also a CCG.
They proved that mate-in-1 and mate-in-2 in Netrunner are weakly NP-hard.
To the most famous card game Magic: The Gathering, Chatterjee and Ibsen-Jensen \cite{magicconphard} showed that simply deciding if a move is legal is coNP-complete in 2016.
And in 2020, Churchill, Biderman, and Herrick \cite{magicturingcomplete} proved that Magic is Turing complete.

More recently, the computational complexity of Hearthstone was concerned.
In 2020, Hoffmann, Lynch, and Winslow \cite{hearthstonenphard} consider the complexity of The Boomsday Project in Hearthstone.
They proved all four puzzle types, Lethal, Mirror, Board Clear, and Survival, in The Boomsday Project to be NP-hard.
The Lethal type is like the mate-in-1 puzzle in Hearthstone, i.e. the problem of whether the friendly hero could kill the enemy hero in this turn.

In this paper, we will prove that the mate-in-$n$ like puzzle in Hearthstone is PSPACE-hard.

\section{Mechanism of Hearthstone}

\textbf{Overview.}
In constructed play, which is the main and classic game mode of Hearthstone, the player does battle with an opponent, each using decks constructed from their own card collections.
If the Health of a player's hero is ever reduced to zero or less, that player loses.
Decks must normally contain exactly 30 cards. 
However, during the game, cards may add or remove from players' decks, and the constraint of 30 cards only applies while building decks, not in the middle of a game. 
Games generally begin with cards in a deck being randomly shuffled.
The game is turn-based.
In each turn, the player gains Mana Crystals, up to 10, and draws a card from the top of his deck. 
Players use Mana to pay for cards and hero powers.
Each card has a Mana cost value (a number in the upper left corner).
The cards themselves often have text which specifies additional abilities and rules that may make game-play differ from the typical behaviour.
There are five types of cards: Minion, Spell, Weapon, Hero, and Location.
In this paper, we just use Minion, Spell, and Weapon cards.

\textbf{Minion Cards.} 
Each minion card has an Attack value (a number in the bottom left corner) and a Health value (a number in the bottom right corner).
Playing a minion card places that minion (with the Attack value and Health value) onto the battlefield, and this process is called ``summoning''.
Once summoned the minion will stay on the battlefield until it is destroyed.
Minions can be destroyed by reducing their Health to zero, or by using destroy effects cards.
Each player cannot have more than 7 minions on the battlefield at a time.
Once 7 friendly minions are on the battlefield, the player will not be able to summon further minions. 
When summoning a minion, the player can choose where to place it.
However, once a minion has been placed it cannot be moved or replaced.

In each turn, each minion can attack one target (an enemy minion or hero) once.
But note that, the minions which are summoned in the current turn can not attack, and they can attack in the next turn.
In combat between minions, each minion will take damage equal to the Attack of the other. 
This damage is dealt to both targets simultaneously.
If either minion's health is reduced to zero or less they will be destroyed and removed from the battlefield.
With combat involving heroes, the direction of attack is important. 
If a hero attacks a minion, both targets will take damage equal to the Attack of the other, but if a minion attacks a hero, only the hero will take damage.
If a minion causes a hero's health to reach zero, the hero is destroyed, ending the game.
Minions can also be affected by other cards (such as buff or healing cards).
For example, ``Give a minion +2/+3'' means increasing the minion's Attack by 2 and increasing the minion's Health by 3.
Minions can also have special abilities (described on the cards) that affect gameplay while they are on the battlefield.

Several specific minion types exist.
A minion's type does not directly affect the behaviour of the minion but allows it to be affected by certain type-specific cards.
Most minions do not have a type, but where present minion type is stated at the bottom of the card.
In this paper, we just use two minion types, Beast and Demon.

\textbf{Spell Cards.}
Spell cards are cards that usually can be played to trigger a one-time effect or ability, described in the card's text. 
Casting a spell consumes the card.
While spells with the Secret or Quest ability are delayed until activated by specific events, and we do not use Secret or Quest spells in this paper. 
Spell cards do not have Attack or Health attributes, only a Mana cost, shown in the top left corner.
Spell cards provide a range of functions, including damage-dealing, removal of minions, providing useful enchantments, drawing cards, and restoring Health.

\textbf{Weapon Cards.} 
Weapon cards can be equipped with heroes. 
Each weapon has an Attack value (a number in the bottom left corner) and a Durability value (a number in the bottom right corner). 
Each hero can only equip one weapon at a time, and equipping a new weapon will destroy the old one. 
While a weapon is equipped, the hero's Attack will be increased by an amount equal to the weapon's Attack, and attacking while a weapon is equipped will reduce its Durability by 1. 
When a weapon's Durability is reduced to zero, the weapon is destroyed.
Usually a hero could only attack once in a turn.
And note that, when a hero attacks a minion, the hero will take damage equal to the Attack of the minion.
In this paper, we only use one weapon card, Light's Justice.

\textbf{Keywords.}
There are a lot of keywords in Hearthstone.
Here we briefly explain 5 keywords used in this paper, including Taunt, Battlecry, Deathrattle, Charge, and Freeze.
Minions with Taunt serve to protect their allies by forcing the enemy to deal with them first, preventing them from attacking other friendly minions or the hero until the minions with Taunt are destroyed.
Battlecry is an ability where the stated effect activates when the card with the Battlecry is played directly from the hand.
Especially, when the minion with Battlecry is summoned, the corresponding effect activates.
Deathrattle is an ability where the stated effect activates when the minion or weapon with Deathrattle is destroyed.
Charge is an ability allowing a minion to attack in the same turn it is summoned.
Freeze is a mechanic that marks a minion or hero as Frozen.
Frozen characters miss their next possible attack.
After this, the Frozen status is removed.

\textbf{Generalizing.}
In the game of Hearthstone, all elements are bounded. 
For example, each turn has a time limit of 75 seconds (plus animations), games are limited to 89 turns, decks are limited to 60 cards, hands to 10 cards, and boards to 14 minions (7 for each player). 
However, the computational complexity of problems is considered as the problem size grows to infinity.
So we have to consider generalizing the game of Hearthstone.
Hoffmann, Lynch, and Winslow \cite{hearthstonenphard} introduced three generalizations of Hearthstone, arbitrarily large boards, hands, and decks. 
In this paper, we mainly consider the generalization of arbitrarily large decks, i.e. players’ decks may have arbitrarily many cards (beyond the 60 permitted in the game), and turns of the game are not bounded certainly, because this generalization seemed to be most natural.
In the actual game, cards are drawn uniformly at random from the deck. 
However, in this paper, we consider the perfect information edition of Hearthstone, i.e. each player knows the entire configuration, including the opponent's hand and card ordering of bilateral decks, and no cards with random elements are used in our reduction. 
The problem of a configuration of Hearthstone is to determine whether the current player could win the game, i.e. can the player reduce the health of the opponent's hero to zero?
Note that, the lethal problem in The Boomsday Project Lab is to ask whether the player could win in the current turn, which is like the mate-in-1 problem in Chess.
While in this paper, we consider the mate-in-$n$ like puzzle for Hearthstone.

\textbf{Others.}
In Hearthstone, new cards are released regularly in sets. 
Only the core set and most recent sets are allowed in the Standard format. 
However, all cards are allowed in the Wild format. 
So we only discuss the computational complexity of Wild format in this paper.
All cards used in our reduction were released before August 2017.
In other words, we will show that perfect information Hearthstone is PSPACE-hard even just using cards in set Basic, Classic, Curse of Naxxramas, Goblins vs Gnomes, Blackrock Mountain, The Grand Tournament, League of Explorers, Whispers of the Old Gods, One Night in Karazhan, Mean Streets of Gadgetzan, Journey to Un'Goro, and Knights of the Frozen Throne.

\section{Complexity of Hearthstone}

\subsection{Partition Game}

To prove the PSPACE-hardness of Hearthstone, we need a PSPACE-hard or PSPACE-complete problem for reduction.
Wolfe \cite{goendgames} proved partition game to be PSPACE-complete in his paper in which the computational complexity of Go endgames is shown to be PSPACE-hard.
We briefly describe the partition game here:

Instance: A collection of $2n$ non-negative integers $x_i$ and $y_i$ ($1 \leq i \leq n$), and a target integer $T$.

Question: Players Left and Right alternate selecting numbers for inclusion in the set $S$. 
Firstly, player Left chooses $x_1$ or $y_1$, and then player Right chooses $x_2$ or $y_2$, and then Left chooses $x_3$ or $y_3$, and so forth. 
At the end of the game, $S$ will have $n$ elements.
Left wins if and only if the elements sum to exactly $T$.

For example, the instance of a partition game is
$$
x_1 =1, \ y_1 =2, \ x_2 =4, \ y_2 =3, \ x_3 =5, \ y_3 =6, \ x_4 =8, \ y_4 =8, \ T=18.
$$

Suppose that, in this instance, players Left and Right choose 
$$x_1 =1, \ y_2 =3, \ y_3 =6, \ x_4 =8$$
alternately, then the sum of these four numbers is exactly equal to the target $18$, so Left wins the partition game.

In the next two subsections, we will demonstrate how to simulate the above instance of the partition game in Hearthstone.

\subsection{Overview of the Reduction}

To prove the PSPACE-hardness of Hearthstone, we should reduce the partition game to Hearthstone.
So we should show that, for each instance of partition game, there is an instance of Hearthstone such that players Left has a forced win in the partition game if and only if the friendly hero has a forced win in Hearthstone.
We will construct a Hearthstone instance for the partition game instance described in the last subsection to illustrate the reduction.
This instance contains a lot of elements including heroes, minions on the battlefield, cards in hand, and cards in decks.
And we will explain that the friendly hero has a forced win in Hearthstone if and only if player Left has a forced win in the partition game.

In the instance of Hearthstone, two heroes are both Rogue, and each hero has only 1 Health and the weapon Light's Justice.
Since the heroes have only 1 Health, they can not attack minions, or they will die immediately.
So the weapon Light's Justice is ``useless'' in the reduction, while Light's Justice cards are padding cards in the instance.
And the hero power of Rogue (gain a 1/2 weapon) will be never used in the reduction.
In the beginning, two players have no cards in hand, and cards in the decks and the order of these cards will be described in detail in the next subsection.

For each $x_i$ or $y_i$ in the instance of partition game, we will set a minion (Floating Watcher or Gahz'rilla) with $10 x_i +2$ or $10 y_i +2$ Attack.
For the target $T$, we set a minion (Leper Gnome) with Taunt and $10T+2 n + 8$ Health.
In each turn of Hearthstone, the player should summon two minions and cast buff cards on them until their Attack reaches $10x_i$ or $10 y_i$.
Note that each player just draws one card at the beginning of a turn, so the player may not have enough buff cards to cast.
To overcome this, we use Gadgetzan Auctioneer's ability (``Whenever you cast a spell, draw a card'') to achieve drawing cards continuously in one turn.
So we set one Gadgetzan Auctioneer on the battlefield for each player.

Then the player should choose to reserve one of these two minions and destroy another, which is corresponding to choosing from $x_i$ or $y_i$ in the partition game.
The reserved minion should attack the Taunt Leper Gnome with $10T+2n+8$ Health in this turn (or in the next turn).

The elements sum to exactly $T$ in the partition game if and only if the Taunt Leper Gnome is destroyed in the proper turn, and the player Left wins in the partition game if and only if the friendly hero wins in Hearthstone.
Note that, in our reduction, each player's hero has only 1 health, thus suppose that the enemy Leper Gnome is destroyed in an earlier time, the Leper Gnome's ability (``Deathrattle: Deal 2 damage to the enemy hero'') will cause friendly hero's health to zero.
So friendly hero must kill the Leper Gnome in the proper turn, and in that turn friendly hero may restore Health, otherwise she will lose the game.

The initial battlefield of the instance is shown in Table \ref{initial_battlefield}.
The function of the Leper Gnome and two Gadgetzan Auctioneer is explained in the last three paragraphs.
The ability ofMistress of Mixtures (``Deathrattle: Restore 4 Health to each hero'') is restoring Health to heroes, and it should attack enemy minions and die in the proper turn, so that friendly minions could kill the enemy Leper Gnome, and the friendly hero will not die.
Wee Spellstopper's ability (``Adjacent minions can't be targeted by spells'') limits the targets of buff spell cards.
Note that, at the beginning, all friendly minions are Frozen, thus they could not attack in the first turn.

\begin{table}[htbp]\footnotesize 
	\centering
	\caption{Initial battlefield.}
	\label{initial_battlefield}       
	\begin{tabular}{ccccccc}
		\hline\noalign{\smallskip}
		Enemy & & Leper & Wee & Wee & Gadgetzan &   \\
		Minions & & Gnome & Spellstopper & Spellstopper & Auctioneer &  \\
		\noalign{\smallskip}\hline\noalign{\smallskip}
		Attack & & sufficiently large & 2 & 2 & 4 &  \\
		\noalign{\smallskip}
		MAX Health & & $10T+2n+8$ & 5 & 5 & 4 &  \\
		\noalign{\smallskip}
		Health & & $10T+2n+8$ & 5 & 5 & 4 &  \\
		\noalign{\smallskip}
		Status & & Taunt &  &  & &  \\
		\noalign{\smallskip}\hline\noalign{\smallskip}
		\noalign{\smallskip}\hline\noalign{\smallskip}
		Friendly & Wee & Wee & Mistress of & Wee & Wee & Gadgetzan \\
		Minions & Spellstopper & Spellstopper & Mixtures & Spellstopper & Spellstopper & Auctioneer \\
		\noalign{\smallskip}\hline\noalign{\smallskip}
		Attack & 2 & 2 & 2 & 2 & 2 & sufficiently large \\
		\noalign{\smallskip}
		MAX Health & 5 & 5 & 2 & 5 & 5 & 4 \\
		\noalign{\smallskip}
		Health & 5 & 5 & 2 & 5 & 5 & 4 \\
		\noalign{\smallskip}
		Status & Frozen & Frozen & Frozen & Frozen & Frozen & Taunt, Frozen  \\
		\noalign{\smallskip}\hline
	\end{tabular}
\end{table}

Not all cards can be included in a constructed deck, most notably each hero has a class and decks can only contain neutral cards or cards from their class.
While we use cards from several classes, so we need to show that it is possible that cards of any class occur in the game.
Fortunately, Pilfer which is a Rogue card could generate any cards from another class, which enables that cards of any class could occur in players' hands.
And Dead Man's Hand could shuffle hand cards into decks, which enables any sequence of cards could occur in decks.
Moreover, Players could use Lorewalker Cho to generate enough Pilfer and Dead Man's Hand cards.
Players could use Acherus Veteran to increase the Attack of minions, and Crazed Alchemist could swap the Attack and Health of a minion.
So it is possible to increase the Attack and Health of minions to any needed values.
Sunfury Protector and Frost Nova could give Taunt and Frozen status to minions respectively.
Deathwing cards are used to eliminate redundant cards in hand.

Here we explain how a specific card which is denoted by X occurs in decks.
The main tactic is to use the Rogue class card, Pilfer, to generate Dead Man's Hand.
Then players could use the Dead Man's Hand to shuffle all cards in hand into decks.
Combining the ability of Lorewalker Cho, any number of cards of all classes could be generated and shuffled into decks.
Now we describe the tactic in detail.
Suppose that, there are two Lorewalker Cho, two Dead Man's Hand, one Pilfer, one Light's Justice, and one Deathwing in friendly hero hand, and there are no minions on the battlefield.
At first, the friendly hero summons one Lorewalker Cho.
Then she cast one Dead Man's Hand.
And then she cast the Pilfer to gain one X card.
At last, she cast another Dead Man's Hand.
Note that, in this process, two Lorewalker Cho, one Dead Man's Hand, one Pilfer, two Light's Justice, two Deathwing, and one X are shuffled in her deck.
Among these cards, the X should be shuffled in a specific position.
While two Lorewalker Cho, two Light's Justice, and one Deathwing are used to ensure that another specific card is generated and shuffled in the deck, these cards should be shuffled on the top of the deck.
And due to the limit of Mana, the friendly hero could not summon the Deathwing in this turn, and she can only summon it in the next turn to clear all cards in hand, so one Light's Justice should be on the top of the deck.
While a Dead Man's Hand, a Pilfer, and a Deathwing are redundant, these cards should be shuffled at the bottom of the deck.
Note that, in this process, due to the ability of Lorewalker Cho, the enemy hero gains two Dead Man's Hand and a Pilfer in hand, so he also could shuffle specific cards in his deck similarly.
In addition, cards generated by Pilfer are class cards, while we could use a similar tactic to shuffle any neutral card to any position of the deck.

Thus it is possible that any cards of any order occur in the decks.
In this way, we could set cards of a specific order in decks to achieve the initial battlefield, so the initial battlefield and needed decks in the reduction are both possible in Hearthstone.

\subsection{A Detailed Example}

In this subsection, we use a detailed example to show how to simulate the instance of the partition game in Hearthstone.
In the instance of Hearthstone, two heroes are both Rogue, and each hero has only 1 Health and the weapon Light's Justice, and they have no cards in hand.
The friendly hero is corresponding to player Left, and the enemy hero is corresponding to player Right.
We use one turn in Hearthstone to simulate one choice in the partition game, and cards of each turn are listed in tables.

Note that, in the instance of partition game, $x_1 =1, y_1 =2, x_2 =4, y_2 =3, x_3 =5, y_3 =6, x_4 =8, y_4 =8, T=18.$
So Health of the Leper Gnome is $10T+2n+8=10 \times 18 + 2 \times 4+8 = 196$.
And Attack of the Leper Gnome and the friendly Gadgetzan Auctioneer is sufficiently large, such as 1000.
There are 5 turns in the instance of Hearthstone, while the first 4 turns are used to simulate 4 choices in the instance of partition, and the last turn is used to simulate verifying whether the sum is equal to $T$ or not.

\begin{table}[htbp]\footnotesize 
	\centering
	\caption{Friendly deck for the first turn.}
	\label{deck_first_turn}       
	\begin{tabular}{lll}
		\hline\noalign{\smallskip}
		Cards & Function description & Partition game    \\
		\noalign{\smallskip}\hline\noalign{\smallskip}
		Innervate$\times 9$, Arcane Intellect, & get Mana and draw cards & \\
		\noalign{\smallskip}
		Floating Watcher, & get a Demon minion & \\
		\noalign{\smallskip}
		Light's Justice, & padding & \\
		\noalign{\smallskip}
		Innervate$\times 4$, & get Mana & \\
		\noalign{\smallskip}
		Demonfuse$\times 2$, & increase Floating Watcher's Attack to 10 & $x_1 =1$ \\
		\noalign{\smallskip}
		Innervate$\times 7$, Arcane Intellect, & get Mana and draw cards & \\
		\noalign{\smallskip}
		Charge, Shadow Word: Death, & kill Floating Watcher or give it Charge & choose $x_1$ or not \\
		\noalign{\smallskip}
		Light's Justice, & padding & \\
		\noalign{\smallskip}
		Innervate$\times 10$, Arcane Intellect, & get Mana and draw cards & \\
		\noalign{\smallskip}
		Gahz'rilla, & get a Beast minion & \\
		\noalign{\smallskip}
		Light's Justice, & padding & \\
		\noalign{\smallskip}
		Innervate$\times 7$, & get Mana & \\
		\noalign{\smallskip}
		[Mark of Y'Shaarj, Light's Justice]$\times$2, & increase Gahz'rilla's Attack to 10 & \\
		\noalign{\smallskip}
		Innervate$\times 9$, & get Mana & \\
		\noalign{\smallskip}
		Blessed Champion, & increase Gahz'rilla's Attack to 20 & $y_1 = 2$ \\
		\noalign{\smallskip}
		Light's Justice, & padding & \\
		\noalign{\smallskip}
		Innervate$\times 9$, Arcane Intellect, & get Mana and draw cards & \\
		\noalign{\smallskip}
		Floating Watcher, & get a Demon minion for the next turn & \\
		\noalign{\smallskip}
		Light's Justice, & padding & \\
		\noalign{\smallskip}
		Innervate$\times 6$, & get Mana & \\
		\noalign{\smallskip}
		Demonfuse, & get a 7 Attack Floating Watcher & \\
		\noalign{\smallskip}
		Frost Nova, & Freeze enemy minions & \\
		\noalign{\smallskip}
		Light's Justice. & padding & \\
		\noalign{\smallskip}\hline
	\end{tabular}
\end{table}

\textbf{The first turn.}
This turn is used to simulate choosing between $x_1$ and $y_1$ in the partition game.
The friendly deck for the first turn is shown in Table \ref{deck_first_turn}.
Note that all friendly minions are frozen, and the friendly hero could not attack minions since she has only 1 Health, so the friendly hero can not kill the enemy hero in this turn.
Suppose that the friendly hero does not cast Frost Nova in this turn, the enemy Leper Gnome could attack and kill the friendly Taunt Gadgetzan Auctioneer in the next turn.
Since the Attack of the Gadgetzan Auctioneer is large enough to destroy the Leper Gnome, the ability of the Leper Gnome will kill the friendly hero immediately.
Thus friendly hero has to draw and cast the Frost Nova in this turn.

The construction of the deck ensures that the friendly hero has enough Mana to play all cards in the deck, and the ability of the Gadgetzan Auctioneer enables her to draw cards continuously.
Drawing cards is the core of the reduction.
Since we set Light's Justice as padding in particular positions of the decks, drawing fewer cards results in that the player could not draw the Frost Nova, and she will be killed in the next turn.
The friendly hero should equip all Light's Justice once she has enough Mana in order to prevent the number of cards in hand from reaching the upper limit.

At the beginning of this turn, the friendly hero draws an Innervate, and she has to cast Innervate.
Due to the ability of the friendly Gadgetzan Auctioneer, she can draw another Innervate, and she has to cast Innervate continuously until drawing an Arcane Intellect.
After casting Arcane Intellect, she has a Floating Watcher, a Light's Justice, and an Innervate in hand.
Then she should summon the Floating Watcher, equip the Light's Justice, and cast Innervate.
After casting 4 Innervate, she has a Demonfuse in hand.
Note that, since the target of Demonfuse should be a Demon minion, the Floating Watcher needs to be on the battlefield at this moment.
Moreover, the Floating Watcher should not be adjacent to those Wee Spellstopper, otherwise, it could not be the target of any spell cards, so the Floating Watcher must be summoned in the rightmost slot.
After casting 2 Demonfuse, the Attack of the Floating Watcher becomes 10, which is corresponding to $x_1=1$ in the instance of the partition game.

Then friendly hero draws and casts 7 Innervate and an Arcane Intellect so that she has a Charge, a Shadow Word: Death, and a Light's Justice in hand.
Note that due to the ability of Wee Spellstopper, only the Floating Watcher could be the target of Charge or Shadow Word: Death.
Then we discuss three situations.

(i) Suppose that Shadow Word: Death is cast on the Floating Watcher.
This is corresponding to that $x_1$ is not chosen in the partition game.
And the Charge must be cast on the following Gahz'rilla.

(ii) Suppose that Charge is cast on the Floating Watcher, and it attacks the enemy Leper Gnome.
Then since the Attack of the Leper Gnome is very large, the Floating Watcher will be destroyed, and the Health of the Leper Gnome will reduce by $10 x_1+2=12$.
This is corresponding to that $x_1$ is chosen in the partition game.

(iii) Suppose that Charge is cast on the Floating Watcher and it does not attack.
Then friendly hero must cast the Shadow Word: Death on the Floating Watcher, otherwise she will have no slot to summon the following Gahz'rilla.
However, since the Charge and Shadow Word: Death are both cast on the Floating Watcher, the Gahz'rilla will never be destroyed in this turn, which will result in the second Floating Watcher can not be summoned.

After the Floating Watcher is destroyed (no matter whether due to attacking or the Shadow Word: Death), the friendly hero draws and casts 10 Innervate and an Arcane Intellect, and then she draws a Gahz'rilla.
She must summon the Gahz'rilla, and note that the Gahz'rilla must be summoned in the rightmost slot, otherwise, the following Mark of Y'Shaarj can not be cast on a Beast, which results in that friendly hero drawing fewer cards, and she can never draw the Frost Nova.
Then, still due to the ability of Wee Spellstopper, only the Gahz'rilla could be the target of two Mark of Y'Shaarj and a Blessed Champion.
After these three spells are cast, the friendly hero has a Gahz'rilla with 20 Attack which is corresponding to $y_1 =2$ in the instance of the partition game.

Then friendly hero has a Charge or a Shadow Word: Death in hand.
To draw more cards, she must cast the spell on the Gahz'rilla.
If the spell is a Shadow Word: Death, the Gahz'rilla will be destroyed, which is corresponding to that $y_1=2$ is not chosen in the partition game.
If the spell is a Charge, the Gahz'rilla should attack the enemy Taunt Leper Gnome.
Since the Attack of the Leper Gnome is very large, the Floating Watcher will be destroyed, and the Health of the Leper Gnome will reduce by $10 y_1+2=22$.
This is corresponding to that $y_1$ is chosen in the partition game.
Suppose that the Gahz'rilla does not attack, the following Floating Watcher could not be summoned due to the limit of slots, which will result in no minions could be the target of the following Demonfuse.

Then friendly hero draws and casts 9 Innervate and an Arcane Intellect so that she has a Floating Watcher, a Light's Justice, and a Demonfuse in hand.
After summoning the second Floating Watcher on the rightmost and casting Demonfuse on the Floating Watcher, the friendly hero has a Demon minion with 7 Attack on the battlefield.
At the end of this turn, the friendly hero casts the Frost Nova to Freeze all enemy minions and equips all Light's Justice.

In the instance of the partition game, the player Left chooses $x_1=1$, so in the instance of Hearthstone, the 12 Attack Floating Watcher attacks the Taunt Leper Gnome, so the Health of Leper Gnome reduces to $196-12=184$.

\begin{table}[htbp]\footnotesize 
	\centering
	\caption{Enemy deck for the second turn.}
	\label{deck_second_turn}       
	\begin{tabular}{lll}
		\hline\noalign{\smallskip}
		Cards & Function description & Partition game    \\
		\noalign{\smallskip}\hline\noalign{\smallskip}
		Innervate$\times 2$, & get Mana & \\
		\noalign{\smallskip}
		Demonfuse, & increase Floating Watcher's Attack to 10 & \\
		\noalign{\smallskip}
		Innervate$\times 10$, & get Mana & \\
		\noalign{\smallskip}
		Blessed Champion$\times 2$, & increase Floating Watcher's Attack to 40 & $x_2 = 4$ \\
		\noalign{\smallskip}
		Innervate$\times 6$, & get Mana & \\
		\noalign{\smallskip}
		Mortal Coil$\times 6$, & deal damage to Floating Watcher & \\
		\noalign{\smallskip}
		Innervate$\times 10$, & get Mana & \\
		\noalign{\smallskip}
		Mind Control, & control friendly Floating Watcher & \\
		\noalign{\smallskip}
		Innervate$\times 10$, Arcane Intellect, & get Mana and draw cards & \\
		\noalign{\smallskip}
		Gahz'rilla, & get a Beast minion & \\
		\noalign{\smallskip}
		Light's Justice, & padding & \\
		\noalign{\smallskip}
		Innervate$\times 7$, & get Mana & \\
		\noalign{\smallskip}
		[Mark of Y'Shaarj, Light's Justice]$\times$2, & increase Gahz'rilla's Attack to 10 & \\
		\noalign{\smallskip}
		Backstab, & increase Gahz'rilla's Attack to 20 & \\
		\noalign{\smallskip}
		Innervate$\times 10$, & get Mana & \\
		\noalign{\smallskip}
		[Mark of Y'Shaarj, Light's Justice]$\times$5, & increase Gahz'rilla's Attack to 30 & $y_2 = 3$ \\
		\noalign{\smallskip}
		Innervate$\times 7$, & get Mana & \\
		\noalign{\smallskip}
		Shadow Word: Death, & kill Floating Watcher or Gahz'rilla & choose $x_2$ or $y_2$ \\
		\noalign{\smallskip}
		Innervate$\times 4$, & get Mana & \\
		\noalign{\smallskip}
		Frost Nova, & Freeze friendly minions & \\
		\noalign{\smallskip}
		Light's Justice. & padding & \\
		\noalign{\smallskip}\hline
	\end{tabular}
\end{table}

\textbf{The second turn.}
This turn is used to simulate choosing between $x_2$ and $y_2$ in the partition game.
The enemy deck for the second turn is shown in Table \ref{deck_second_turn}.
Similarly, all enemy minions are frozen, and the enemy hero could not attack minions since he has only 1 Health.
Suppose that the enemy hero does not cast Frost Nova in this turn, then in the next turn, the friendly Mistress of Mixtures could attack enemy Leper Gnome and be destroyed, which restores 4 Health to each hero.
Note that, the Health of the Leper Gnome is at least 8 (see Table \ref{initial_battlefield}), so it could not be destroyed by the Mistress of Mixtures.
Then the friendly hero has 5 Health, so the friendly Gadgetzan Auctioneer could kill the enemy Taunt Leper Gnome.
Leper Gnome's Deathrattle only deals 2 damage to the friendly hero, so she will not die.
Then the friendly hero and minions could attack and kill the enemy hero.  
Thus the enemy hero also has to draw and cast the Frost Nova in this turn.

Then the enemy hero has to cast Innervate to draw the Demonfuse and 2 Blessed Champion.
Due to the ability of Wee Spellstopper, the Demonfuse, 2 Blessed Champion, and the following spells must be cast on the friendly Floating Watcher.
So that the Attack of the Floating Watcher reaches 40, which is corresponding to $x_2=4$ in the instance of the partition game.

Then the enemy hero has to cast 6 Mortal Coil cards to deal 6 damage to the Floating Watcher, which makes it to be a damaged minion.
This ensures that the following Backstab will not be cast on the Floating Watcher, since only undamaged minions could be the target of Backstab.
Note that the Mortal Coil cards can not destroy the Floating Watcher, since it has 10 Health.
Then enemy hero has to cast the Mind Control, so that the Floating Watcher becomes the enemy's rightmost minion, i.e. it is not adjacent to any Wee Spellstopper.

Then enemy hero has to cast Innervate and the Arcane Intellect.
And he should summon the Gahz'rilla and cast 2 Mark of Y'Shaarj on it, which increases Gahz'rilla's Attack to 10.
Note that the Floating Watcher also can be the target of the Mark of Y'Shaarj, but casting on it will result in drawing fewer cards since the Floating Watcher is not a Beast minion.
For the same reason, the Mark of Y'Shaarj should not be cast on other minions, which means that Gahz'rilla is not adjacent to any Wee Spellstopper.

Then enemy hero has a Backstab in hand.
Note that the Floating Watcher is damaged, and other minions (except the Gahz'rilla) are all adjacent to Wee Spellstopper, so the Backstab has to be cast on the Gahz'rilla.
Due to the ability of Gahz'rilla, the Backstab will double the Attack of the Gahz'rilla.
Adding the following 5 Mark of Y'Shaarj, the Attack of the Gahz'rilla will reach 30, which is corresponding to $y_2=3$ in the instance of the partition game.

Then enemy hero has a Shadow Word: Death in hand.
Note that, on the battlefield, only the Floating Watcher and the Gahz'rilla are not adjacent to any Wee Spellstopper, so the enemy hero has to choose one of these two minions to cast the Shadow Word: Death.
In the instance of the partition game, the player Right chooses $y_2=3$ instead of $x_2=4$.
So in the instance of Hearthstone, the enemy hero casts the Shadow Word: Death on the Floating Watcher with 40 Attack and destroys it.

At the end of this turn, the enemy hero casts the Frost Nova to Freeze all friendly minions and equips all Light's Justice, and he has a 30 Attack Gahz'rilla on the battlefield.

\begin{table}[htbp]\footnotesize 
	\centering
	\caption{Friendly deck for the third turn.}
	\label{deck_third_turn}       
	\begin{tabular}{lll}
		\hline\noalign{\smallskip}
		Cards & Function description & Partition game    \\
		\noalign{\smallskip}\hline\noalign{\smallskip}
		Innervate$\times 10$, & get Mana & \\
		\noalign{\smallskip}
		Mind Control, & control enemy minion & \\
		\noalign{\smallskip}
		Innervate$\times 7$,  & get Mana & \\
		\noalign{\smallskip}
		Charge, & give the minion Charge & choose $x_2$ or $y_2$ \\
		\noalign{\smallskip}
		Novice Engineer, Mortal Coil, & ensure the controlled minion attacks & \\
		\noalign{\smallskip}
		Light's Justice, & padding & \\
		\noalign{\smallskip}
		Innervate$\times 9$, Arcane Intellect, & get Mana and draw cards & \\
		\noalign{\smallskip}
		Floating Watcher, & get a Demon minion & \\
		\noalign{\smallskip}
		Light's Justice, & padding & \\
		\noalign{\smallskip}
		Innervate$\times 4$, & get Mana & \\
		\noalign{\smallskip}
		Demonfuse$\times 2$, & increase Floating Watcher's Attack to 10 & \\
		\noalign{\smallskip}
		Innervate$\times 10$, & get Mana & \\
		\noalign{\smallskip}
		Blessed Champion$\times 2$, & increase Floating Watcher's Attack to 40 & \\
		\noalign{\smallskip}
		Innervate$\times 10$, Mark of Y'Shaarj$\times 5$, & increase Floating Watcher's Attack to 50 & $x_3 =5$  \\
		\noalign{\smallskip}
		Innervate$\times 7$, Arcane Intellect, & get Mana and draw cards & \\
		\noalign{\smallskip}
		Charge, Shadow Word: Death, & kill Floating Watcher or give it Charge & choose $x_3$ or not \\
		\noalign{\smallskip}
		Light's Justice, & padding & \\
		\noalign{\smallskip}
		Innervate$\times 10$, Arcane Intellect, & get Mana and draw cards & \\
		\noalign{\smallskip}
		Gahz'rilla, & get a Beast minion & \\
		\noalign{\smallskip}
		Light's Justice, & padding & \\
		\noalign{\smallskip}
		Innervate$\times 7$, & get Mana & \\
		\noalign{\smallskip}
		[Mark of Y'Shaarj, Light's Justice]$\times$2, & increase Gahz'rilla's Attack to 10 & \\
		\noalign{\smallskip}
		Innervate$\times 5$, & get Mana & \\
		\noalign{\smallskip}
		Blessed Champion, & increase Gahz'rilla's Attack to 20 & \\
		\noalign{\smallskip}
		Innervate$\times 10$, & get Mana & \\
		\noalign{\smallskip}
		[Mark of Y'Shaarj, Light's Justice]$\times$5, & increase Gahz'rilla's Attack to 30 & \\
		\noalign{\smallskip}
		Innervate$\times 10$, & get Mana & \\
		\noalign{\smallskip}
		Blessed Champion, & increase Gahz'rilla's Attack to 60 & $y_3 = 6$ \\
		\noalign{\smallskip}
		Innervate$\times 4$, & get Mana to cast Charge or Shadow Word & \\
		\noalign{\smallskip}
		Light's Justice, & padding & \\
		\noalign{\smallskip}
		Innervate$\times 9$, Arcane Intellect, & get Mana and draw cards & \\
		\noalign{\smallskip}
		Floating Watcher, & get a Demon minion for the next turn & \\
		\noalign{\smallskip}
		Light's Justice, & padding & \\
		\noalign{\smallskip}
		Innervate$\times 6$, & get Mana & \\
		\noalign{\smallskip}
		Demonfuse, & get a 7 Attack Floating Watcher & \\
		\noalign{\smallskip}
		Frost Nova, & Freeze enemy minions & \\
		\noalign{\smallskip}
		Light's Justice. & padding & \\
		\noalign{\smallskip}\hline
	\end{tabular}
\end{table}

\textbf{The third turn.}
This turn is used to simulate choosing between $x_3$ and $y_3$ in the partition game.
The friendly deck for the third turn is shown in Table \ref{deck_third_turn}.
Similarly, the friendly hero also has to draw and cast the Frost Nova in this turn.

Friendly hero cast Innervate to draw and cast the Mind Control.
Note that, due to the ability of Wee Spellstopper, only the enemy Floating Watcher or Gahz'rilla could be the target of Mind Control, so the friendly hero has to cast the Mind Control and control this enemy minion.
And this controlled minion becomes the friendly rightmost minion.

Then friendly hero casts Innervate and the Charge.
Also due to Wee Spellstopper, the Charge must be cast on the controlled minion.
Note that, at this time, the friendly hero has only a Novice Engineer in hand, so she has to summon Novice Engineer to draw more cards.
However, there are 7 friendly minions on the battlefield, to summon Novice Engineer, the controlled minion must attack the enemy Taunt Leper Gnome, which simulates choosing $x_2$ or $y_2$ in the partition game.
In the instance, $y_2 = 3$ is chosen, so the 32 Attack Gahz'rilla attacks the Leper Gnome.
Then the Health of Leper Gnome reduces to $184-32=152$.

Now friendly hero has a Mortal Coil in hand, while the next card in the deck is a Light's Justice.
This means that the Mortal Coil should destroy a minion to trigger the ability to draw cards.
And note that, there is only one minion with 1 Health on the battlefield, the Novice Engineer.
Therefore, the Novice Engineer must be summoned on the rightmost to avoid any Wee Spellstopper, and the friendly hero must cast the Mortal Coil on the Novice Engineer to kill it and draw one additional card.

The following process is similar to the first turn.
Friendly hero summons a Floating Watcher, and uses Demonfuse, Mark of Y'Shaarj, and Blessed Champion to increase its Attack to 50, which is corresponding to $x_3$ in the partition game.
Then friendly hero chooses to cast the Charge or Shadow Word: Death, which simulates choosing $x_3$ or not.
And then the Floating Watcher will attack the enemy Taunt Leper Gnome or be destroyed by Shadow Word: Death.
Friendly hero summons a Gahz'rilla, and uses Mark of Y'Shaarj and Blessed Champion to increase its Attack to 60, which is corresponding to $y_3$ in the partition game.
At the end of this turn, the friendly hero has a Floating Watcher with 7 Attack on the battlefield, and casts the Frost Nova to Freeze all enemy minions.

In the instance of the partition game, the player Left chooses $y_3=6 $, so in the instance of Hearthstone, the 62 Attack Gahz'rilla attacks the Taunt Leper Gnome, so the Health of Leper Gnome reduces to $152-62=90$.

\begin{table}[htbp]\footnotesize 
	\centering
	\caption{Enemy deck for the fourth turn.}
	\label{deck_fourth_turn}       
	\begin{tabular}{lll}
		\hline\noalign{\smallskip}
		Cards & Function description & Partition game    \\
		\noalign{\smallskip}\hline\noalign{\smallskip}
		Innervate$\times 7$, & get Mana & \\
		\noalign{\smallskip}
		Demonfuse, Blessed Champion, & increase Floating Watcher's Attack to 20 & \\
		\noalign{\smallskip}
		Innervate$\times 10$, & get Mana & \\
		\noalign{\smallskip}
		Blessed Champion$\times 2$, & increase Floating Watcher's Attack to 80 & $x_4 = 8$ \\
		\noalign{\smallskip}
		Innervate$\times 6$, & get Mana & \\
		\noalign{\smallskip}
		Mortal Coil$\times 6$, & deal 6 damage to Floating Watcher & \\
		\noalign{\smallskip}
		Innervate$\times 10$, & get Mana & \\
		\noalign{\smallskip}
		Mind Control, & control friendly Floating Watcher & \\
		\noalign{\smallskip}
		Innervate$\times 10$, Arcane Intellect, & get Mana and draw cards & \\
		\noalign{\smallskip}
		Gahz'rilla, & get a Beast minion & \\
		\noalign{\smallskip}
		Light's Justice, & padding & \\
		\noalign{\smallskip}
		Innervate$\times 7$, & get Mana & \\
		\noalign{\smallskip}
		[Mark of Y'Shaarj, Light's Justice]$\times$2, & increase Gahz'rilla's Attack to 10 & \\
		\noalign{\smallskip}
		Innervate$\times 8$, & get Mana & \\
		\noalign{\smallskip}
		Backstab, & increase Gahz'rilla's Attack to 20 & \\
		\noalign{\smallskip}
		Flash Heal, & heal Gahz'rilla & \\
		\noalign{\smallskip}
		Backstab, & increase Gahz'rilla's Attack to 40 & \\
		\noalign{\smallskip}
		Flash Heal, & heal Gahz'rilla & \\
		\noalign{\smallskip}
		Backstab, & increase Gahz'rilla's Attack to 80 & $y_4 = 8$ \\
		\noalign{\smallskip}
		Shadow Word: Death, & kill Floating Watcher or Gahz'rilla & choose $x_4$ or $y_4$ \\
		\noalign{\smallskip}
		Frost Nova, & Freeze friendly minions & \\
		\noalign{\smallskip}
		Light's Justice. & padding & \\
		\noalign{\smallskip}\hline
	\end{tabular}
\end{table}

\textbf{The fourth turn.}
This turn is used to simulate choosing between $x_4$ and $y_4$ in the partition game.
The enemy deck for the fourth turn is shown in Table \ref{deck_fourth_turn}.
Similarly, the enemy hero also has to draw and cast the Frost Nova in this turn.

The process in this turn is similar to the process in the second turn.
The enemy hero summons a Floating Watcher, and he uses a Demonfuse and Blessed Champion to increase its Attack to 80, which is corresponding to $x_4 = 8$ in the partition game.
Then the enemy hero cast 6 Mortal Coil on the Floating Watcher, which deals 6 damage to it, and he casts the Mind Control on the Floating Watcher.
And then enemy hero summons a Gahz'rilla, and he uses Mark of Y'Shaarj and Backstab to increase its Attack to 80, which is corresponding to $y_4 = 8$ in the partition game.
Enemy hero casts the Shadow Word: Death on the Floating Watcher or Gahz'rilla, which is corresponding to choosing from $x_4$ or $y_4$ in the partition game.
At the end of this turn, the enemy hero casts the Frost Nova to Freeze all friendly minions.

Note that, there are two Flash Heal in the deck, and we explain why these Flash Heal must be cast on the Gahz'rilla here.
Due to the ability of Wee Spellstopper, only the Floating Watcher and Gahz'rilla could be the target of spell cards.
Suppose that the Flash Heal is cast on the Floating Watcher, note that the Floating Watcher takes 6 damage by Mortal Coil, so it is still damaged even if a Flash Heal is cast on it.
This results in that the Backstab could not be cast since there are no undamaged minions on the battlefield.
Suppose that the Flash Heal is cast on heroes, there are no undamaged minions on the battlefield either.
In this situation, the enemy hero can not draw the Frost Nova in this turn, and in the next turn, the friendly hero and minions could kill the enemy Taunt Leper Gnome and the enemy hero.

In the instance of partition game, the player Right chooses $x_4=8$, so in the instance of Hearthstone, the 80 Attack Gahz'rilla is destroyed by the Shadow Word: Death, and the 80 Attack Floating Watcher is on the battlefield.

\begin{table}[htbp]\footnotesize 
	\centering
	\caption{Friendly deck for the fifth turn.}
	\label{deck_fifth_turn}       
	\begin{tabular}{lll}
		\hline\noalign{\smallskip}
		Cards & Function description & Partition game    \\
		\noalign{\smallskip}\hline\noalign{\smallskip}
		Innervate$\times 10$, & get Mana & \\
		\noalign{\smallskip}
		Mind Control, & control enemy minion & \\
		\noalign{\smallskip}
		Innervate$\times 7$,  & get Mana & \\
		\noalign{\smallskip}
		Charge, & give the minion Charge & choose $x_4$ or $y_4$ \\
		\noalign{\smallskip}
		Novice Engineer, Mortal Coil, & ensure the controlled minion attacks & \\
		\noalign{\smallskip}
		Light's Justice, & padding & \\
		\noalign{\smallskip}
		Innervate$\times 10$, Arcane Intellect, & get Mana and draw cards & \\
		\noalign{\smallskip}
		Gahz'rilla, & get a minion & \\
		\noalign{\smallskip}
		Light's Justice, & padding & \\
		\noalign{\smallskip}
		Innervate$\times 6$, & get Mana & \\
		\noalign{\smallskip}
		Flash Heal, & heal friendly hero & \\
		\noalign{\smallskip}
		Charge, & give Gahz'rilla Charge & verify sum is equal to $T$ or not \\
		\noalign{\smallskip}
		Light's Justice. & padding & \\
		\noalign{\smallskip}\hline
	\end{tabular}
\end{table}

\textbf{The fifth turn.}
This turn is used to simulate verifying whether the sum is equal to $T$ or not in the partition game.
The enemy deck for the fourth turn is shown in Table \ref{deck_fifth_turn}.

The first half part of the deck is similar to the third turn.
Friendly hero casts the Mind Control and Charge on the enemy minion.
Then the controlled minion must attack the enemy Taunt Leper Gnome, which simulates choosing $x_4$ or $y_4$ in the partition game.
In the instance of partition game, the player Right chooses $x_4=8$, so the 82 Attack Floating Watcher attacks the Taunt Leper Gnome, so the Health of Leper Gnome reduces to $90-82=8$.

So far, the process of simulating two players' choices in the partition game is over.
Then friendly hero summons the Novice Engineer, and cast the Mortal Coil on it.
The following deck simulates verifying whether the sum of integers is equal to $T$ in the partition game.
The friendly hero should summon a Gahz'rilla, then she casts the Flash Heal on herself and then casts the Charge on the Gahz'rilla.
Then the Gahz'rilla should attack the enemy Taunt Leper Gnome.
And friendly hero should attack and kill the enemy hero if the Leper Gnome is destroyed by the Gahz'rilla.

We consider the remaining Health of the enemy Taunt Leper Gnome.
Note that, in the beginning, the Leper Gnome has $10T+2n+8$ Health.
According to the decks, for each $x_i$ ($y_i$) that is chosen in the partition game, there is a minion with $10 x_i +2$ ($10 y_i +2$) Attack which attacked the Leper Gnome.
Thus, before the last friendly Gahz'rilla attacks, the remaining Health of the Leper Gnome is either zero (i.e. the Leper Gnome is already destroyed) or $10k+8$ ($k=0,1,2, \dots$).
Here we discuss three situations according to the remaining Health of the Leper Gnome.

(i) Suppose that the Leper Gnome is already destroyed.
This is corresponding to that the sum of elements chosen by two players is larger than the target $T$ in the partition game.
Then due to the ability of the Leper Gnome, the friendly hero must have been dead.

(ii) Suppose that the Health of the Leper Gnome is $10k+8$ ($k=1,2,3, \dots$).
This is corresponding to that the sum of elements chosen by two players is smaller than the target $T$ in the partition game.
Since friendly Gahz'rilla can not kill the Leper Gnome in this turn, and friendly hero has no Frost Nova cards, the enemy hero and minions could kill friendly Taunt Gadgetzan Auctioneer and hero in the next turn even if the Flash Heal is cast on friendly hero.

(iii) Suppose that the Health of the Leper Gnome is exactly 8, just like in the instance.
This is corresponding to that the sum of elements chosen by two players is equal to the target $T$ in the partition game.
Then the friendly could cast the Flash Heal on herself to restore 5 Health, and the friendly Gahz'rilla could just right destroy the enemy Taunt Leper Gnome, and the friendly hero could attack and kill the enemy hero in this turn.

At the bottom of the enemy deck, we set a Light's Justice in order to avoid Fatigue damage.

Therefore, we show that the friendly hero could kill the enemy hero in this turn if and only if the sum of chosen elements is equal to the target $T$, i.e. player Left has a forced win in the instance of the partition game if and only if the friendly hero has a forced win in the instance of Hearthstone.
Note that, Blessed Champion and Gahz'rilla could double the Attack of minions in Hearthstone.
So the reduction could be constructed in polynomial time even if the number in the partition game is exponential.
Since the partition game is PSPACE-complete, we obtain the following result:

\begin{theorem}
	Perfect information Hearthstone is PSPACE-hard.
\end{theorem}

\section{Comments}

To the instance of partition game, we construct an instance of Hearthstone, such that we show that mate-in-$n$, i.e. whether the friendly hero could kill the enemy hero in $n$ turns, in perfect information Hearthstone is PSPACE-hard.

It would be very interesting to know whether the game remains hard (or is harder) with random elements.
It is also worth considering the computational complexity of Hearthstone generalizations of arbitrarily large battlefield or hands.

If we want to show the EXPTIME-hardness or even EXPSPACE-hardness of Hearthstone, we may use the formula games introduced in \cite{expcproblem} and \cite{expspacecproblem}.
In these formula games, the number of turns is exponential instead of polynomial.
So, to simulate the formula games, the number of turns in Hearthstone should also be exponential.
However the number of cards in decks in an instance of Hearthstone is polynomial, and this means the cards which could manipulate decks must be used in the reduction.
However, there are usually random units in these cards, so it is difficult to construct reduction.

\section*{Acknowledgements}

We would like to thank Tian Bai, Zhaobo Han, Chaohao Pan, and Chengyang Qian for their valuable comments.
We especially thank Jayson Lynch for his important comments and suggestions.


\bibliographystyle{plain}
\bibliography{ref}

\section*{Appendix}

We list all cards used in the reduction in alphabet order here.
All pictures are from Hearthstone Wiki (https://hearthstone.fandom.com/wiki/Hearthstone\_Wiki).

\begin{figure}[htbp]
	\centering
	\begin{minipage}[htbp]{0.32\linewidth}
		\centering
		\includegraphics[width=1 \linewidth]{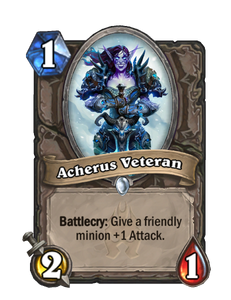}
	\end{minipage}
	\begin{minipage}[htbp]{0.32\linewidth}
		\centering
		\includegraphics[width=1 \linewidth]{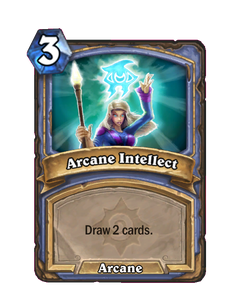}
	\end{minipage}
	\begin{minipage}[htbp]{0.32\linewidth}
		\centering
		\includegraphics[width=1 \linewidth]{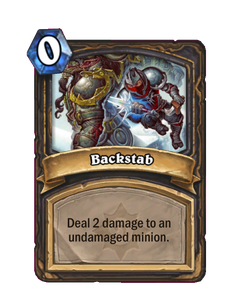}
	\end{minipage}
\end{figure}

\begin{figure}[htbp]
	\centering
	\begin{minipage}[htbp]{0.32\linewidth}
		\centering
		\includegraphics[width=1 \linewidth]{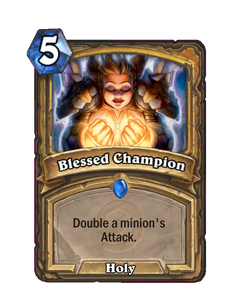}
	\end{minipage}
	\begin{minipage}[htbp]{0.32\linewidth}
		\centering
		\includegraphics[width=1 \linewidth]{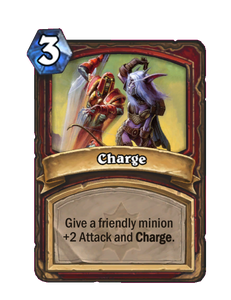}
	\end{minipage}
	\begin{minipage}[htbp]{0.32\linewidth}
		\centering
		\includegraphics[width=1 \linewidth]{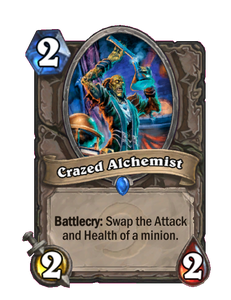}
	\end{minipage}
\end{figure}

\begin{figure}[htbp]
	\centering
	\begin{minipage}[htbp]{0.32\linewidth}
		\centering
		\includegraphics[width=1 \linewidth]{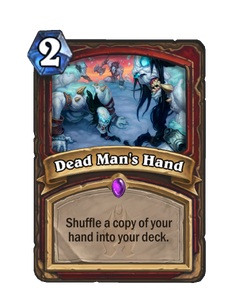}
	\end{minipage}
	\begin{minipage}[htbp]{0.32\linewidth}
		\centering
		\includegraphics[width=1 \linewidth]{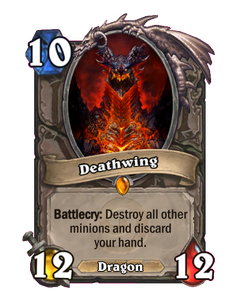}
	\end{minipage}
	\begin{minipage}[htbp]{0.32\linewidth}
		\centering
		\includegraphics[width=1 \linewidth]{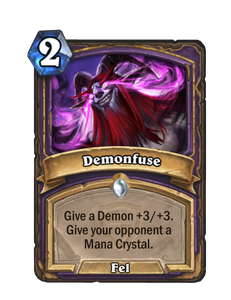}
	\end{minipage}
\end{figure}

\begin{figure}[htbp]
	\centering
	\begin{minipage}[htbp]{0.32\linewidth}
		\centering
		\includegraphics[width=1 \linewidth]{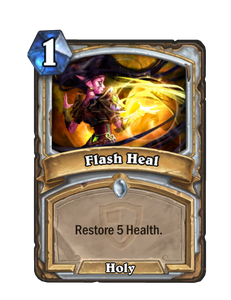}
	\end{minipage}
	\begin{minipage}[htbp]{0.32\linewidth}
		\centering
		\includegraphics[width=1 \linewidth]{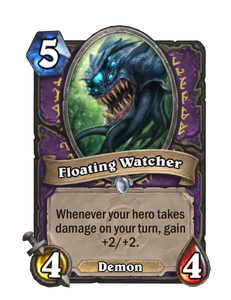}
	\end{minipage}
	\begin{minipage}[htbp]{0.32\linewidth}
		\centering
		\includegraphics[width=1 \linewidth]{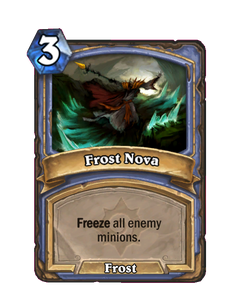}
	\end{minipage}
\end{figure}

\begin{figure}[htbp]
	\centering
	\begin{minipage}[htbp]{0.32\linewidth}
		\centering
		\includegraphics[width=1 \linewidth]{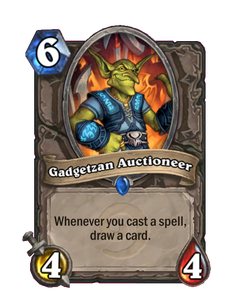}
	\end{minipage}
	\begin{minipage}[htbp]{0.32\linewidth}
		\centering
		\includegraphics[width=1 \linewidth]{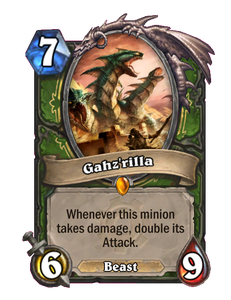}
	\end{minipage}
	\begin{minipage}[htbp]{0.32\linewidth}
		\centering
		\includegraphics[width=1 \linewidth]{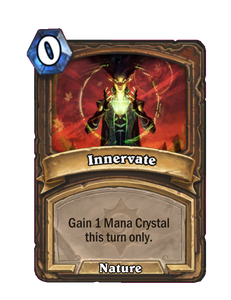}
	\end{minipage}
\end{figure}

\begin{figure}[htbp]
	\centering
	\begin{minipage}[htbp]{0.32\linewidth}
		\centering
		\includegraphics[width=1 \linewidth]{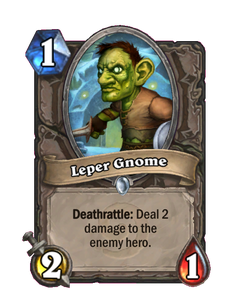}
	\end{minipage}
	\begin{minipage}[htbp]{0.32\linewidth}
		\centering
		\includegraphics[width=1 \linewidth]{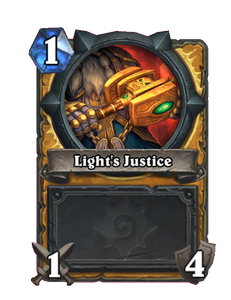}
	\end{minipage}
	\begin{minipage}[htbp]{0.32\linewidth}
		\centering
		\includegraphics[width=1 \linewidth]{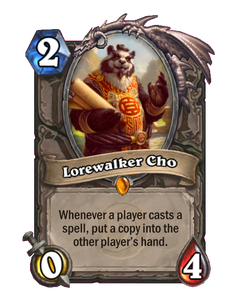}
	\end{minipage}
\end{figure}

\begin{figure}[htbp]
	\centering
	\begin{minipage}[htbp]{0.32\linewidth}
		\centering
		\includegraphics[width=1 \linewidth]{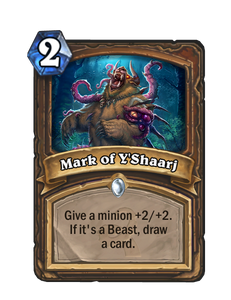}
	\end{minipage}
	\begin{minipage}[htbp]{0.32\linewidth}
		\centering
		\includegraphics[width=1 \linewidth]{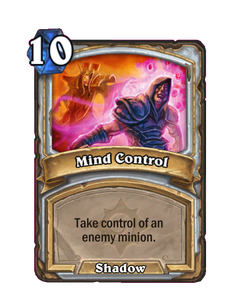}
	\end{minipage}
	\begin{minipage}[htbp]{0.32\linewidth}
		\centering
		\includegraphics[width=1 \linewidth]{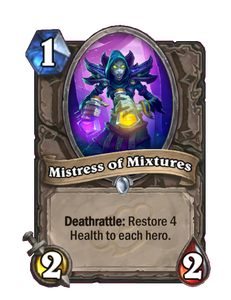}
	\end{minipage}
\end{figure}

\begin{figure}[htbp]
	\centering
	\begin{minipage}[htbp]{0.32\linewidth}
		\centering
		\includegraphics[width=1 \linewidth]{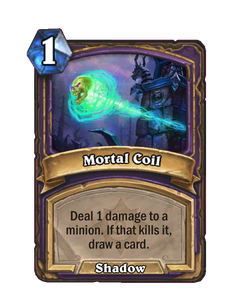}
	\end{minipage}
	\begin{minipage}[htbp]{0.32\linewidth}
		\centering
		\includegraphics[width=1 \linewidth]{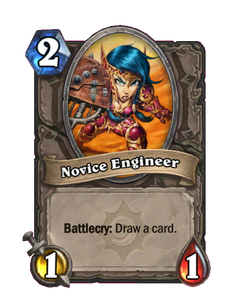}
	\end{minipage}
	\begin{minipage}[htbp]{0.32\linewidth}
		\centering
		\includegraphics[width=1 \linewidth]{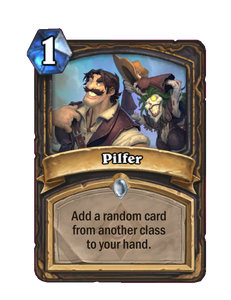}
	\end{minipage}
\end{figure}

\begin{figure}[htbp]
	\centering
	\begin{minipage}[htbp]{0.32\linewidth}
		\centering
		\includegraphics[width=1 \linewidth]{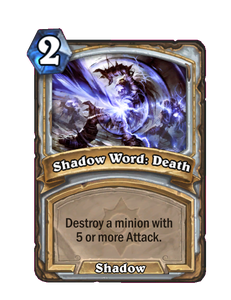}
	\end{minipage}
	\begin{minipage}[htbp]{0.32\linewidth}
		\centering
		\includegraphics[width=1 \linewidth]{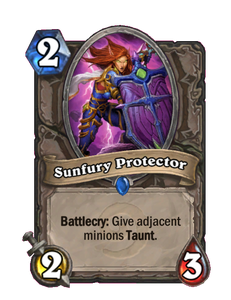}
	\end{minipage}
	\begin{minipage}[htbp]{0.32\linewidth}
		\centering
		\includegraphics[width=1 \linewidth]{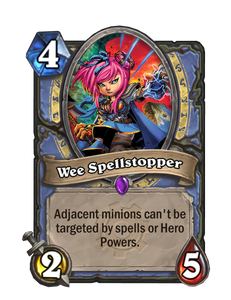}
	\end{minipage}
\end{figure}

\end{document}